\documentclass[a4paper]{report}
\usepackage[utf8]{inputenc}
\usepackage[T1]{fontenc}
\usepackage{RJournal}
\usepackage{amsmath,amssymb,array}
\usepackage{booktabs}

\begin{document}

\sectionhead{Contributed research article}
\volume{XX}
\volnumber{YY}
\year{20ZZ}
\month{AAAA}

\begin{article}
\renewcommand{\backref}[1]{}\pagestyle{plain}

\title{A Unifying Framework for Parallel and Distributed Processing in R using Futures}

\author{by Henrik Bengtsson}

\maketitle

\abstract{
A \dfn{future} is a programming construct designed for concurrent and
asynchronous evaluation of code, making it particularly useful for
parallel processing. The \pkg{future} package implements
the \dfn{Future API} for programming with futures in R.  This minimal
API provides sufficient constructs for implementing parallel versions
of well-established, high-level map-reduce APIs.  The future ecosystem
supports exception handling, output and condition relaying, parallel
random number generation, and automatic identification of globals
lowering the threshold to parallelize code.  The \dfn{Future API}
bridges parallel frontends with parallel backends, following the
philosophy that end-users are the ones who choose the parallel backend
while the developer focuses on what to parallelize.  A variety of
backends exist, and third-party contributions meeting the
specifications, which ensure that the same code works on all backends,
are automatically supported.  The future framework solves several
problems not addressed by other parallel frameworks in R.
}

\section{Introduction}
\label{introduction}

Parallel processing can be used to speed up computationally intensive
tasks.  As the size of these tasks and access to more CPU cores tend
to grow over time, so does the demand for parallel-processing
solutions.  In R, there exist several frameworks for running code in
parallel, many dating back more than a
decade~\citep{Schmidberger2009}.  R gained built-in support via
the \pkg{parallel} package in version 2.14.0 (2011), which to date
probably provides the most, either directly or indirectly, commonly
used solutions.  For an overview of current parallel techniques
available to R developers, see \citet{Eddelbuettel2020} and the
\href{https://CRAN.R-project.org/view=HighPerformanceComputing}{\emph{High-Performance and Parallel Computing with R}} CRAN Task View.

The options for parallelizing \emph{computations} in R can be grouped
broadly into those that can be used to parallelize R code, such as
what the \pkg{parallel} package provides, and those that are used to
parallelize native code, such as C, C++, and Fortran, and are often not
specific to R itself.  For example, multi-threaded processing is an
efficient parallelization technique which operates at the core of the
operating system and the CPU and allows for updating shared memory in
parallel and more, which is not available at the R level.  In contrast,
parallelization at the R level takes place at a higher level with a
coarser type of parallelization, which we refer to as \dfn{multi-process}
parallelization.
In addition to parallel computations, there are also efforts in R for
working with \emph{parallel data structures}, e.g.,
\CRANpkg{sparklyr}~\citep{CRAN:sparklyr} and the
\dfn{Programming with Big Data in R} (pbdR) project~\citep{Schmidt2017}.
By pre-distributing data and storing them on, or near, parallel
workers, the overhead from passing data on-the-fly in parallel
processing can be decreased, resulting in an overall faster processing
time but also lower and more fine-tuned memory requirements.
This article proposes a solution for \emph{parallelizing computations
at the R level}.

The \CRANpkg{future} package~\citep{CRAN:future} aims to provide a
unifying, generic, minimal application protocol interface (API) to
facilitate the most common types of parallel processing in R,
especially the \dfn{manager-worker} strategy where an R process
delegates tasks to other R processes. It builds upon the concepts
of \dfn{futures} \citep{HewittBaker_1977}
and \dfn{promises} \citep{FriedmanWise_1978, Hibbard_1976} - concepts
that are well suited for a functional language such as R.
To better understand how it fits in among and relates to existing
parallelization solutions in R\footnote{Although the concept of
futures could also apply to C, C++, and Fortran parallelization, the
future framework targets parallelization at the R level and does not
provide an implementation for native code.}, let us revisit the two
most well-known solutions - packages \pkg{parallel}
and \CRANpkg{foreach}.

The \pkg{parallel} package has a set of functions for calling
functions and expressions in parallel across one or more concurrent R
processes. The most well-known functions for this
are \code{mclapply()} and \code{parLapply()}, which mimic the behavior
of the map-reduce\footnote{We use the term ``map-reduce'' as it is
used in functional programming. The \dfn{MapReduce} method by
\citet{DeanGhemawat2004} was inspired by this term but they are not
equivalent.}
function \code{lapply()} in the \pkg{base} package.  Below is an
example showing them calling a ``slow'' function on each element in a
vector using two parallel workers.  First, to do this through
sequentially processing, we can use \code{lapply()}:

\begin{example}
xs <- 1:10
y <- lapply(xs, function(x) {
  slow_fcn(x)
})
\end{example}
To do the same in parallel using two \emph{forked} parallel processes, we
can use:

\begin{example}
library(parallel)
xs <- 1:10
y <- mclapply(xs, function(x) {
  slow_fcn(x)
}, mc.cores = 2)
\end{example}
Alternatively, to run it in parallel using two R parallel processes
running in the \emph{background}, we can do:

\begin{example}
library(parallel)
workers <- makeCluster(2)
clusterExport(workers, "slow_fcn")
xs <- 1:10
y <- parLapply(workers, xs, function(x) {
  slow_fcn(x)
})
\end{example}
These functions, which originate from legacy packages
\CRANpkg{multicore}~(2009-2014, \citet{CRAN:multicore}) and
\CRANpkg{snow}~(since 2003, \citet{CRAN:snow}), are designed for
specific parallelization frameworks. The \code{mclapply()} set of
functions relies on process \emph{forking} by the operating system,
which makes them particularly easy to use.  This is because each
worker automatically inherits the setup and all of the content of the
main R process' workspace, making it straightforward to replace a
sequential
\code{lapply()} call with a parallel \code{mclapply()} call. This
has made it popular among Linux and macOS developers. On MS Windows,
where R does not support forked processing, \code{mclapply()} falls
back to using \code{lapply()} internally.

The \code{parLapply()} set of functions, which all operating systems
support, rely on a cluster of R background workers for
parallelization. It works by the main R process and the workers
exchanging tasks and results over a communication channel. The
default and most commonly used type of cluster is SOCK, which MS
Windows also supports, and it communicates via \emph{socket
connections}. Like most other cluster types, SOCK clusters require
developers to manually identify and export packages and global objects
to the workers by calling \code{clusterEvalQ()} and
\code{clusterExport()}, before calling \code{parLapply()}, which
increases the barrier to use them.

\subsection{Mixed responsibilities of developers or end-users}
\label{mixed-responsibilities}

Using either the \code{mclapply()} or the \code{parLapply()} approach
works well when developers and end-users can agree on which framework
to use. Unfortunately, this is not always possible, e.g., R package
developers rarely know who the end-users are and what compute
resources they have. Regardless, developers who wish to support
parallel processing still face the problem of deciding which parallel
framework to target, a decision that often has to be done early in
the development cycle. This means deciding on what \emph{type of
parallelism} to support, e.g., forked processing via \code{mclapply()}
or SOCK clusters via \code{parLapply()}. This decision is critical
because it limits the end-user's options, and any change, later on,
might be expensive because of, for instance, having to rewrite and
retest part of the codebase. A developer who wishes to support
multiple parallel backends has to implement support for each of them
individually and provide the end-user with a mechanism to choose
between them. This approach often results in unwieldy,
hard-to-maintain code of conditional statements with low test
coverage, e.g.,
\begin{example}
if (parallel == "fork") {
  ...
} else if (parallel == "SOCK") {
  ...
} else if (parallel == "MPI") {
  ...
} else {
  ...
}
\end{example}
There is no established standard for doing this, which results in
different packages providing different mechanisms for controlling the
parallelization method, if at all.

Functions like \code{parLapply()} partly address the problem of
supporting multiple parallelization frameworks because they support
various types of parallel cluster backends referred to as ``snow''
clusters (short for \dfn{Simple Network of Workstations} and from
their origin in the \pkg{snow} package), e.g.,
\code{workers\,<-\,makeCluster(4, type\,=\,"FORK")} sets up a cluster that
parallelizes using forked processing, and
\code{workers\,<-\,makeCluster(4, type\,=\,"MPI")} sets up a cluster that
parallelizes via a Message Passing Interface (MPI) framework.  If a
developer uses \code{parLapply()}, they could write their code such
that the end-user can specify what type of snow cluster to use, e.g.,
by respecting what the end-user set
via \code{setDefaultCluster(workers)}.  This provides the end-user
with more, although in practice limited, options on how and where to
execute code in parallel. Unfortunately, it is rather common that the
cluster type is hard-coded inside packages giving end-users little to
no control over the parallelization mechanism, other than possibly the
number of cores to use.

\subsection{Map-reduce parallelization with more control for the end-user}
\label{map-reduce-parallelization-with-more-control-for-the-end-user}

Possibly inspired by the snow-style clusters, the \pkg{foreach}
package~\citep{CRAN:foreach,Kane_etal_2013}, first released in 2009,
addresses the above problem of having to decide on the parallel design
early on by letting the end-user - not the developer - ``register''
what type of parallel backend (``foreach adaptor'') to use when
calling \code{foreach()}. For example, with
\CRANpkg{doMC}~\citep{CRAN:doMC}, one can register a multicore cluster, and
with \CRANpkg{doParallel}~\citep{CRAN:doParallel}, one can register any
type of ``snow'' cluster as in:
\begin{example}
library(foreach)
library(doParallel)
workers <- parallel::makeCluster(2)
registerDoParallel(workers)

xs <- 1:10
y <- foreach(x = xs) 
  slow_fcn(x)
}
\end{example}
We note that the specification of what type of parallel framework and
number of cores to use is separated from the \code{foreach()}
map-reduce construct itself. This gives more control to the end-user
on
\emph{how} and \emph{where} to parallelize, leaving the developer to
focus on \emph{what} to parallelize, which is a design pattern of
great value with important implications on how to design, write, and
maintain parallel code. The large uptake of \pkg{foreach} since it was
first released supports this. As of November 2021, \pkg{foreach} is among
the top-1.0\% most downloaded packages on CRAN, and there are 867
packages on CRAN and Bioconductor that directly depend on it.  Another
advantage of the separation between the map-reduce frontend API and
parallel backend (foreach adaptors) is that new types of parallel
backends can be introduced without the need to make updates to
the \pkg{foreach} package. This has led to third-party developers have
contributed additional foreach adaptors,
e.g., \CRANpkg{doMPI}~\citep{CRAN:doMPI} and
\CRANpkg{doRedis}~\citep{CRAN:doRedis}.

Unfortunately, there is no \emph{exact} specification on what a
foreach adaptor should support and how it should act in certain
situations, which has resulted in adaptors behaving slightly
differently. At their face value, these differences appear innocent
but may cause different outcomes of the same code. In the best case,
these differences result in run-time errors, and in the worst case,
different results. An example of the former is the difference
between \pkg{doMC} on Unix-like systems and \pkg{doParallel} on
Windows. Analogously to \code{mclapply()}, when using \pkg{doMC},
globals and packages are automatically taken care of by the process
forking. In contrast, when using \pkg{doParallel} with ``snow''
clusters, globals and packages need to be identified and explicitly
exported, via additional arguments \code{.export} and \code{.packages}
to \code{foreach()}, to the parallel workers running in the
background.  Thus, a developer that only uses \pkg{doMC} might forget
to test their code with \pkg{doParallel}, where it may fail.  Having
said this, the \pkg{foreach} package does provide a rudimentary
mechanism for automatically identifying and exporting global
variables.  However, it has some limitations, that, in practice, require
the developer to explicitly specify globals to make sure their code
works with more backends.  Some adaptors provide additional options of
their own that are specified as arguments to \code{foreach()}. If the
developer specifies such options, the \code{foreach()} call might not
work with other adaptors.

To develop \code{foreach()} code invariant to the parallel backend
chosen requires a good understanding of how the \pkg{foreach}
framework works and plenty of testing. This lack of strict behavior is
unfortunate and might have grown out of a strategy of wanting to keep
things flexible. On the upside, steps have recently\footnote{See
the \pkg{foreach} issue tracker
at \url{https://github.com/RevolutionAnalytics/foreach}.}  been taken
toward making the behavior more consistent across foreach backends,
suggesting that it is possible to remove several of these weaknesses
through a process of deprecating and removing unwanted side effects
over several release cycles in close collaboration with package
developers currently relying on such backend-specific properties.

\section{The future framework}

The \pkg{future} package defines and implements the \dfn{Future API} -
a minimal, unifying, low-level API for parallel processing, and more.
Contrary to the aforementioned solutions, this package does \emph{not}
offer a parallel map-reduce API per se.  Instead, it focuses on
providing efficient and simple-to-use atomic building blocks that
allow us to implement such higher-level functions elsewhere.

\subsection{Three atomic constructs that unify common parallel design patterns}
\label{low-level-generalized-parallelization-model}

\begin{samepage}
The \dfn{Future API} comprises three fundamental constructs:
\begin{itemize}
\item
  \code{f\,<-\,future(expr)} : evaluates an expression via a
  future (non-blocking, if possible)
\item
  \code{v\,<-\,value(f)} : the value of the future
  expression \code{expr} (blocking until resolved)
\item
  \code{r\,<-\,resolved(f)} : TRUE if future is resolved,
  otherwise FALSE (non-blocking)
\end{itemize}
\end{samepage}
To help understand what a future is, let us start with R's assignment
construct:
\begin{example}
v <- expr
\end{example}
Although it is effectively a single operator, there are two steps in
an assignment: first (i) R evaluates the \dfn{expression} on the
right-hand side (RHS), and then (ii) it assigns the resulting value to
the \dfn{variable} on the left-hand side (LHS).  We can think of
the \dfn{Future API} as giving us full access to these two steps by
rewriting the assignment construct as:
\begin{example}
f <- future(expr)
v <- value(f)
\end{example}
Contrary to the regular assignment construct where the evaluation of
the expression and the assignment of its value are tightly coupled,
the future construct allows us to decouple these steps, which is an
essential property of futures and necessary when doing parallel
processing\footnote{We can find this future-value pattern in several
implementations for parallel processing, including the ones we use in
R. The \code{mcparallel()}-\code{mccollect()} pair of functions
in \pkg{parallel} is one example.  This is why the future-value
abstraction can be mapped onto many of our existing parallel
frameworks in a unified way.}.  Especially, the decoupling allows
us to perform other tasks in-between the step that evaluates the
expression and the step that assigns its value to the target variable.
Here is an example that creates a future that calculates
\code{slow\_fcn(x)} with \code{x} being \code{1}, then reassigns
a different value to \code{x}, and finally gets the value of the
future expression:
\begin{example}
x <- 1
f <- future({
  slow_fcn(x)
})
x <- 2
v <- value(f)
\end{example}
By definition, a future consists of an R expression and any required
objects as they were when the future was created. Above, the recorded
objects are the function \code{slow\_fcn()} and the variable \code{x}
with value \code{1}.  This is why the value of the future is
unaffected by \code{x} getting reassigned a new value after the
future is created but before the value is collected.

We have yet to explain how futures are resolved, that is, how the
future expression is evaluated.  This is the part where futures
naturally extend themselves to asynchronous and parallel processing.
How a future is resolved depends on what \dfn{future backend} is set.
If not specified, the default is to resolve futures sequentially,
which corresponds to setting:
\begin{example}
plan(sequential)
\end{example}
Before we continue, it should be emphasized that the \dfn{Future API}
is designed so that a program using it gives the same results no
matter how and where the futures are resolved, may it be sequentially
on the local machine or in parallel on a remote cluster.  As a
consequence, \emph{the future ecosystem is designed to separate the
responsibilities of the developer from those of the end-user}.  This
allows the developer to focus on the code to be parallelized while the
end-user focuses on how to parallelize.  It is the end-user who
decides on the \code{plan()}.  For example, if they specify:
\begin{example}
plan(multisession)
\end{example}
before calling the above future code, futures will be resolved in
parallel via a SOCK cluster on the local machine similar to what we
used above in the \code{parLapply()} example.  If the end-user instead
specifies \code{plan(multicore)}, futures will be resolved in parallel
in the background via \emph{forked} R processes using the same
framework as \code{mclapply()}.  Importantly, regardless of what
future plan is used, and regardless of whether or not we assigned a
new value to \code{x} after creating the future, the result is always
the same.  Since we, as developers, do not know what backend end-users
will use, we also cannot know \emph{when} a future is resolved.  This
is why we say that ``a future evaluates its expression \emph{at some
point in the future}''.  What we do know is that \code{value()}
returns the value of the future only when it is resolved, and if it is
not resolved, then \code{value()} waits until it is.

Next, let us look at how blocking works by using an example where we
create three futures to be resolved by two parallel workers:
\begin{example}
library(future)
plan(multisession, workers = 2)

xs <- 1:10

f1 <- future({
  slow_fcn(xs[1])
})

f2 <- future({
  slow_fcn(xs[2])
})

f3 <- future({
  slow_fcn(xs[3])
})
\end{example}
Here, the first two futures are created in a non-blocking way because
there are two workers available to resolve them.  However, when we
attempt to create a third future, there are no more workers available.
This causes
\code{future()} to block until one of the workers is available, that is,
until either one or both of the two futures have been resolved.  If
three or more workers are set up, then the third \code{future()} call
would not block.  On the other hand, if \code{plan(sequential)} is
set, then each \code{future()} blocks until the previously created
future has been resolved.
Finally, to retrieve the values of the three futures, we do:
\begin{example}
v1 <- value(f1)
v2 <- value(f2)
v3 <- value(f3)
\end{example}
Although it is common to call \code{value()} on the futures in the
order we created them, we can collect the values in any order, which
is something we will return to later.

\begin{figure}[htbp]
  \begin{center}
    \includegraphics[width=0.69\linewidth]{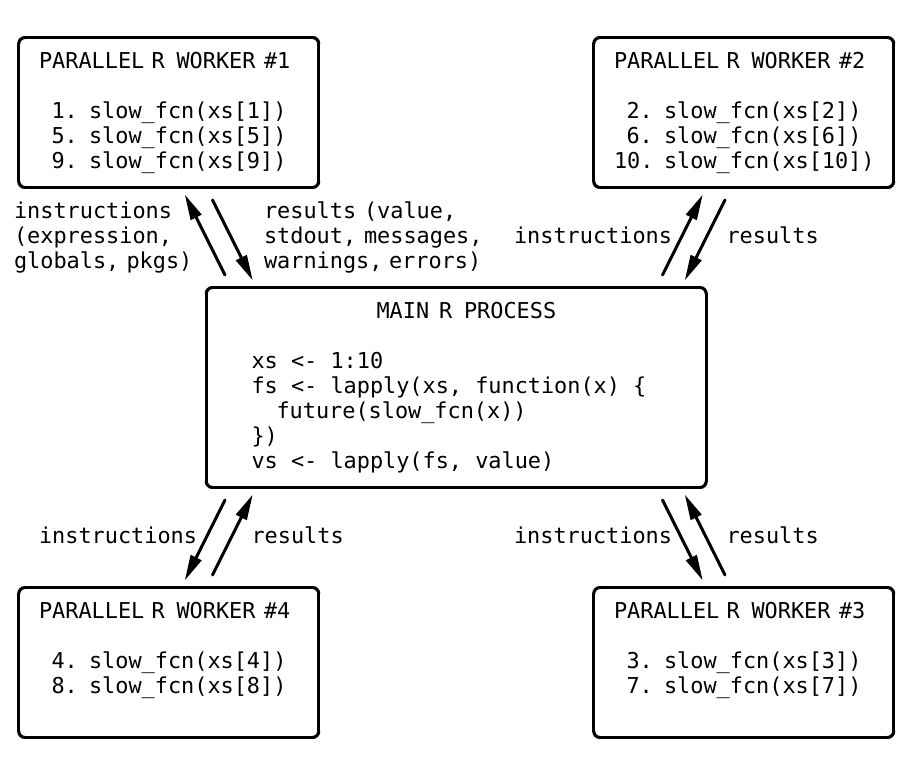}
    \caption{An illustration of parallel processing using futures via
    four R processes running in the background.  Base R
    \code{lapply()} is used to call \code{slow\_fcn()} ten times -
    once per element in \code{xs}.  By calling it via \code{future()},
    each call is distributed out to one of four workers.  If all
    workers are busy, the next future, in turn, will wait for a worker
    to become available.  The results of all futures are collected at
    the end.  Any output, warnings, and errors produced on the workers
    are relayed as-is back on the main R session.  The four workers
    were created using \code{plan(multisession, workers = 4)}.  If
    switching to \code{plan(sequential)}, then all futures are
    resolved sequentially in the main R process.
    Only core Future API functions from the \pkg{future} package were
    used.  Less verbose, map-reduce alternatives are available in the
    high-level future packages such as \pkg{future.apply},
    \pkg{furrr}, and \pkg{doFuture}.}
    \label{figure:future-4workers}
  \end{center}
\end{figure}

Continuing, we can generalize the above to
calculate \code{slow\_fcn()} on each of the elements in \code{xs} via
futures. For this, we can use a regular for-loop to create each of
the \code{length(xs)} futures:
\begin{example}
xs <- 1:10
fs <- list()
for (i in seq_along(xs)) {
  fs[[i]] <- future(slow_fcn(xs[i]))
}
\end{example}
Note how we here have effectively created a \emph{parallel for-loop},
where \code{plan()} controls the amount of parallelization.  To
collect the values of these futures, we can
use\footnote{Here, \code{vs\,<-\,lapply(fs,\,value)} is used for
clarification but we could also have used \code{vs\,<-\,value(fs)}
because \code{value()} is a generic function with implementation also
for lists and other types of containers.}:
\begin{example}
vs <- lapply(fs, value)
\end{example}
Alternatively, to using a for-loop, we can parallelize
using \code{lapply()}:
\begin{example}
xs <- 1:10
fs <- lapply(xs, function(x) {
  future(slow_fcn(x))
})
\end{example}
This is illustrated in Figure~\ref{figure:future-4workers}, where four
background workers created by \code{plan(multisession, workers = 4)}
is used to resolve the futures.
The same idea also applies to other types of map-reduce functions.
This shows how powerful the \dfn{Future API} is; by combining base R
with the two constructs \code{future()} and \code{value()}, we have
created rudimentary\footnote{These solutions process each element in a
separate future, which is suboptimal if the overhead of creating a
future is relatively large compared to the evaluation time. This
overhead can be mitigated by processing elements in chunks, something
that requires more complex code than what is shown in these examples.}
alternatives to \code{mclapply()}, \code{parLapply()}, and
\code{foreach()}.  Indeed, we could reimplemented these
\pkg{parallel} and \pkg{foreach} functions using the \dfn{Future API}.

The \code{resolved()} function queries, in a non-blocking way, whether
or not a future is resolved.  Among other things, this can be used to
collect the value of a subset of resolved futures as soon as possible
without risking to block from collecting the value of a non-resolved
future, which allows additional futures to launch sooner, if they
exist.  This strategy also helps lower the overall latency that comes
from the overhead of collecting values from futures - values that may
contain large objects and are collected from remote machines over a
network with limited bandwidth.  As explained further below,
collecting the value of futures as soon as possible will also lower
the latency of the relay of output and conditions (e.g., warnings and
errors) captured by each future while they evaluate the future
expressions.

In summary, the three constructs of the \dfn{Future API} provide
\emph{the necessary and sufficient} functionality for evaluating
R expressions in parallel, which in turn may be used to construct
higher-level map-reduce functions for parallel processing.  Additional
core features of futures that are useful, or even essential, for
parallel processing are presented next.

\subsection{Exception handling}
\label{error-handling}

To make it as simple as possible to use futures, they are designed to
mimic the behavior of the corresponding code that does not use
futures.  An important part of this design aim is how exception
handling is done.  Any \dfn{error} produced while resolving a future,
that is, evaluating its expression, is captured and relayed as-is in
the main R process each time \code{value()} is called.  This mimics
the behavior of how errors are produced when not using futures.  This
is illustrated by the following two code examples -- with futures:
\begin{example}
x <- "24"
f <- future(log(x))
v <- value(f)
# Error in log(x) : non-numeric argument to mathematical function
\end{example}
and without futures:
\begin{example}
x <- "24"
v <- log(x)
# Error in log(x) : non-numeric argument to mathematical function
\end{example}
As a result, standard mechanisms for condition handling also apply to
errors relayed by futures.  For example, to assign a missing value
to \code{v} whenever there is an error, we can use:
\begin{example}
v <- tryCatch({
  value(f)
}, error = function(e) {
  NA_real_
})
\end{example}

Errors due to extraordinary circumstances, such as terminated R
workers and failed communication, are of a different kind than the above
evaluation errors.  Because of this, they are signaled as errors of
class \dfn{FutureError} so they can be handled specifically, e.g., by
restarting R workers or relaunching the failed future elsewhere
(Section `Future work').

\subsection{Relaying of standard output and conditions (e.g., messages and warnings)}
\label{relaying-of-standard-output-and-conditions-e.g.messages-and-warnings}

Futures capture the standard output (\dfn{stdout}) and then relay it
in the main R process each time \code{value()} is called. Analogously,
all conditions are captured and relayed as-is in the main R process
each time \code{value()} is called. Common conditions relayed this way
are \dfn{message}s and \dfn{warning}s as generated by \code{message()}
and \code{warning()}.  The relaying of errors was discussed in the
previous section.  Relaying of standard output and conditions respects
the order they were captured, except that all of the standard output is
relayed before conditions are relayed in the order they were
signaled. For example,
\begin{example}
x <- c(1:10, NA)
f <- future({
  cat("Hello world\n")
  y <- sum(x, na.rm = TRUE)
  message("The sum of 'x' is ", y)
  if (anyNA(x)) warning("Missing values were omitted", call. = FALSE)
  cat("Bye bye\n")
  y
})
v <- value(f)
# Hello world
# Bye bye
# The sum of 'x' is 55
# Warning message:
# Missing values were omitted
\end{example}
Standard techniques can be used to capture the relayed standard
output, e.g.,
\begin{example}
stdout <- capture.output({
  v <- value(f)
})
# The sum of 'x' is 55
# Warning message:
# Missing values were omitted

stdout
# [1] "Hello world" "Bye bye"
\end{example}
Similarly, \code{withCallingHandlers()}
and \code{globalCallingHandlers()} can be used to capture and handle
the different classes of conditions being relayed.  Note that all of
the above works the same way regardless of what future backend is
used, including when futures are resolved on a remote machine.

Relaying of standard output, messages, warnings, and errors simplifies
any troubleshooting. For example, existing verbose output helps narrow
down the location of errors and warnings, which may reveal unexpected
missing values or vector recycling. Commonly used poor-man debugging,
where temporary debug messages are injected into the code, is also
possible because of this built-in relay mechanism.  Imagine a logging
framework that leverages R's condition framework to signal different
levels of log events and then captures and reports, e.g., to the
terminal or to file.  It will work out of the box when parallelizing
with futures.

Conditions of class \dfn{immediateCondition} are treated specially by
the future framework. They are by design allowed to be relayed as soon
as possible, and not only when \code{value()} is called. For instance,
they may be relayed when calling \code{resolved()}, or even sooner,
depending on the future backend used. Because of this,
\dfn{immediateCondition} conditions are relayed without respecting
the order of other types of conditions captured. This makes them
suitable for signaling, for instance, \emph{progress updates}.  Thus,
such progress conditions can be used to update a progress bar in the
terminal or in a Shiny application while originating from futures
being resolve on remote machines.  See the \CRANpkg{progressr}
package~\citep{CRAN:progressr} for an implementation of this.  Note,
however, that this type of near-live relaying of
\dfn{immediateCondition}s only works for backends that have the means to
communicate these conditions from the worker back to the main R
session, while the worker still processes the future.  When
non-supporting backends are used, these conditions are relayed
together with other captured conditions at the very end when the
future has been resolved.

\emph{Comment:} Contrary to the standard output, due to limitations in
R\footnote{See \url{https://github.com/HenrikBengtsson/Wishlist-for-R/issues/55}},
it is not possible to capture the standard error reliably.  Because of
this, any output to the standard error is silently ignored,
e.g., \code{cat("some~output",\,file\,=\,stderr())}. However, although
output from \code{message()} is sent to the standard error, it is
indeed outputted in the main R processes because it is the message
conditions that are captured and relayed, not the standard error.

\subsection{Globals and packages}
\label{global-variables}

The future framework is designed to make it as simple as possible to
implement parallel code.  Another example of this is the automatic
identification of \dfn{globals} - short for global variables and functions
- that are required for a future expression to be resolved
successfully.  For example, in:
\begin{example}
f <- future({
  slow_fcn(x)
})
\end{example}
the globals of the future expression are \code{slow\_fcn()}
and \code{x}.  By default, \code{future()} will attempt to identify,
locate, and record these globals internally via static code
inspection, such that they are available when the future is resolved.
If one of these globals is part of a package namespace, that is also
recorded.  Because of this, developers rarely need to worry about
globals when programming with futures.
However, occasionally, the future expression is such that it is not
possible to infer all the globals.  For example, the following
produces an error:
\begin{example}
plan(multisession)
k <- 42
f <- future({
  get("k")
})
v <- value(f)
# Error in get("k") : object 'k' not found
\end{example}
This is because code inspection cannot infer that \code{k} is a needed
variable.  In such cases, one can guide the future framework to
identify this missing global by explicitly mentioning it at the top of
the future expression, e.g.,
\begin{example}
f <- future({
  k
  get("k")
})
\end{example}
Alternatively, one can specify it via argument \code{globals} when
creating the future, e.g.,
\begin{example}
f <- future({
  get("k")
}, globals = "k")
\end{example}
See \code{help("future",\,package\,=\,"future")} for all options
available to control which globals to use and how to ignore false
positives.

Internally, the future framework
uses \CRANpkg{globals}~\citep{CRAN:globals}, and
indirectly \CRANpkg{codetools}~\citep{CRAN:codetools}, to identify
globals by walking the abstract syntax tree (AST) of the future
expression in order.  It uses an \emph{optimistic} search strategy to
allow for some false-positive globals to minimize the number of
false-negative globals.  Contrary to false positives, false negatives
cause futures to produce errors similar to the one above.

\subsection{Proper parallel random number generation}
\label{proper-parallel-random-number-generation-rng}

The ability to produce high-quality random numbers is essential for
the validity of many statistical analyses, e.g., bootstrap,
permutation tests, and simulation studies. R has functions at its core
for drawing random numbers from common distributions. This R
functionality is also available to C and Fortran native code. All draw
from the same internal pseudo-random number generator (RNG). Different
kinds of RNGs are available, with
Mersenne-Twister \citep{Matsumoto_Nishimura:1998} being the
default. Like most other RNGs, the Mersenne-Twister RNG is not
designed for concurrent processing - if used in parallel, one risks
producing random numbers that are correlated. Instead, for parallel
processing, the multiple-recursive generator L'Ecuyer-CMRG by
\citet{LEcuyer:1999}, implemented in the \pkg{parallel} package, can be
used to set up multiple RNG streams.
The future ecosystem has built-in support for L'Ecuyer-CMRG at its
core to make it as easy as possible to produce statistically sound and
reproducible random numbers regardless of how and where futures are
resolved, e.g.,
\begin{example}
f <- future(rnorm(3), seed = TRUE)
value(f)
#  [1] -0.02648871 -1.73240257  0.78139056
\end{example}
Above, \code{seed\,=\,TRUE} is used to specify that parallel RNG
streams should be used.  When used, the result will be fully
reproducible regardless of future backend specified and the number of
workers available.  Because \code{seed\,=\,TRUE} can introduce
significant overhead, the default is \code{seed\,=\,FALSE}.  However,
since it is computationally cheap to detect when a future expression
produced random numbers, the future framework will generate an
informative warning when this is used by mistake to help lower the
risk of producing statistically questionable results.  It is possible
to disable this check or to escalate the warning to an error via an R
option.
All higher-level parallelization APIs that build upon futures must
adhere to this parallel-RNG design, e.g., \CRANpkg{future.apply}
and \CRANpkg{furrr}.

\subsection{Future assignment construct}
\label{secFutureAssignmentConstruct}

As an alternative for using \code{future()} and \code{value()},
the \pkg{future} package provides a \dfn{future-assignment
operator}, \code{\%<-\%}, for convenience.  It is designed to mimic
the regular assignment operator, \code{<-}, in R:
\begin{example}
v <- expr
\end{example}
By replacing the above with:
\begin{example}
v 
\end{example}
the RHS expression \code{expr} will be evaluated using a future whose
value is assigned to the LHS variable \code{v} as a
\emph{promise}\footnote{The type of promises that R supports should
not be mistaken for the type of promises as defined by the
\CRANpkg{promises}~\citep{CRAN:promises} package, which, together with
futures, is used for asynchronous processing in Shiny applications.}.
Because the LHS is a promise, the value of the future will not be
assigned to it until we attempt to access the promise.  As soon as we
try to use \code{v}, say,
\begin{example}
y <- sqrt(v) 
\end{example}
the associated promise will call \code{value()} on the underlying
future, while possibly blocking, and at the end assign the collected
result to \code{v}\footnote{The internal call to \code{value()} will
also cause any captured standard output and conditions to be
relayed.}.  From there on, \code{v} is a regular value.  As an
illustration, our introductory example with three futures can be
written as\footnote{I have dropped the curly brackets on the RHS to
make the example tidier. Just like with regular assignment, there is
nothing preventing us from using composite expressions also with
future assignments.}:
\begin{example}
xs <- 1:10
v1 
v2 
v3 
\end{example}
and with, say, \code{plan(multisession)}, these statements will be
processed in parallel.

Special \dfn{infix operators} are available to specify arguments that
otherwise would be passed to the \code{future()} function.  For
example, to set \code{seed = TRUE}, we can use:
\begin{example}
v 
\end{example}
See \code{help("\%<-\%",\,package\,=\,"future")} for other infix
operators.

Regular R assignments can often be replaced by future assignments
as-is.  However, because future assignments rely on promises, and
promises can only be assigned to \dfn{environment}s, including the
working environment, they cannot be used to assign to, for
instance, \dfn{list}s.  As a workaround, one can use a \dfn{list
environment} instead of a \dfn{list}.  They are implemented in
the \CRANpkg{listenv} package~\citep{CRAN:listenv}.  A list
environment is technically an \dfn{environment} that emulates most 
properties of a \emph{list}, including indexing as in:
\begin{example}
xs <- 1:10
vs <- listenv::listenv()
for (i in seq_along(xs)) {
  vs[[i]] 
}
vs <- as.list(vs)
\end{example}

\subsection{Nested parallelism and protection against it}
\label{protection-against-nested-parallelism}

A problem with parallel processing in software stacks like the R
package hierarchy is the risk of overloading the CPU cores due to
nested parallelism. For instance, assume that package \pkg{PkgA} calls
\code{PkgB::estimate()} in parallel using all $N$ cores on the
current machine. Initially, the \code{estimate()} function was
implemented to run sequentially, but, in a recent \pkg{PkgB} release,
it was updated to parallelize internally using all $N$ cores.  Without
built-in protection, this update now risks running $N^2$ parallel
workers when \pkg{PkgA} is used, possibly without the awareness of
either maintainer.

The \pkg{future} package has built-in protection against nested
parallelism. This works by configuring each worker to run in
sequential mode unless nested parallelism is explicitly configured.
This is achieved by setting options and environment variables that are
known to control parallelism in R,
e.g., \code{options(mc.cores\,=\,1)}.  Because of this, if \pkg{PkgA}
and \pkg{PkgB} parallelize using the future framework, the nested
parallelism above will run with a total of $N$ cores, not $N^2$ cores.
This will also be true for non-future code that respects such
settings, e.g., when \pkg{PkgB} uses \code{parallel::mclapply()} with
the default \code{mc.cores} argument.

Nested parallelism can be configured by the end-user
via \code{plan()}.  For example, to use two workers for the first
layer of parallelization and three for the second, use:
\begin{example}
plan(list(
  tweak(multisession, workers = 2),
  tweak(multisession, workers = 3)
))
\end{example}
This will run at most $2 \times 3 = 6$ tasks in parallel on the local
machine.  Any nested parallelism beyond these two layers will be
processed in sequential mode. That is, \code{plan(sequential)} is
implicit if not specified.  When argument \code{workers} is not
specified, it defaults to \code{parallelly::availableCores()}, which
respect a large number of environment variables and R options
specifying the number of cores.  Because of this, and due to the
built-in protection against nested parallelism, using
\code{plan(list(multisession,\linebreak[0]\,multisession))} effectively equals using
\code{plan(list(multisession,\linebreak[0]\,sequential))}.

A more common scenario of nested parallelism is when we submit tasks
to a job scheduler on a compute cluster where each job is allowed to
run on multiple cores allotted by the scheduler.  As clarified later,
this may be configured as:
\begin{example}
plan(list(
  future.batchtools::batchtools_sge,
  multisession
))
\end{example}
where the default \code{workers = availableCores()} assures that the
number of multisession workers used respects what the scheduler
assigns to each job.

\subsection{Future backends}
\label{future-backends}

In addition to implementing the \dfn{Future API}, the \pkg{future} package
also implements a set of future backends that are based on
the \pkg{parallel} package.  If no backend is specified, the default
is:
\begin{example}
plan(sequential)
\end{example}
which makes all futures to be resolved sequentially in the current R
session.
To resolve futures in parallel on a SOCK cluster on the local machine,
use one of:
\begin{example}
plan(multisession) ## defaults to workers = availableCores()
plan(multisession, workers = 4)
\end{example}
Similarly, to resolving futures in parallel on the local machine via
\emph{forked} processing, use one of:
\begin{example}
plan(multicore) ## defaults to workers = availableCores()
plan(multicore, workers = 4)
\end{example}
To resolve futures via \emph{any} type of ``snow'' cluster, use
the \code{cluster} backend.  For example, to use a traditional SOCK
cluster or an MPI cluster, use either of:
\begin{example}
workers <- parallel::makeCluster(4)
plan(cluster, workers = workers)

workers <- parallel::makeCluster(4, type = "MPI")
plan(cluster, workers = workers)
\end{example}
To use a SOCK cluster with two remote workers, use:
\begin{example}
plan(cluster, workers = c("n1.remote.org", "n2.remote.org"))
\end{example}
which is short for:
\begin{example}
workers <- parallelly::makeClusterPSOCK(c("n1.remote.org", "n2.remote.org"))
plan(cluster, workers = workers)
\end{example}
This works as long as there is password-less SSH access to these
remote machines and they have R installed.  Contrary to
\code{parallel::makePSOCKcluster()},
\code{parallelly::makeClusterPSOCK()} uses reverse-tunneling
techniques, which avoids having to configure inward-facing
port-forwarding in firewalls, something that requires administrative
rights.

\subsubsection{Third-party future backends}

Besides these built-in future backends, other R packages available on
CRAN implement additional backends.  As long as these backends conform
to the \dfn{Future API} specifications, as discussed in
Section~'Validation', they can be used as alternatives to the built-in
backends.
For example, the \CRANpkg{future.callr}
package~\citep{CRAN:future.callr} implements a future backend that
resolves futures in parallel on the local machine via R
processes\footnote{The callr backend performs similarly to the
PSOCK-based multisession backend. However, in contrast to the latter,
it does not rely on socket connections, which on MS Windows may
require administrative rights on the machine's firewall in order to
allow the R process to communicate on certain ports. Moreover, on
machines with a large number of cores, PSOCK clusters are limited to
125 parallel workers because that is the maximum number of
connections R can have open simultaneously.}, orchestrated by
the \CRANpkg{callr}~\citep{CRAN:callr} package, e.g.,
\begin{example}
plan(future.callr::callr)  ## defaults to workers = availableCores()
plan(future.callr::callr, workers = 4)
\end{example}
Another example
is \CRANpkg{future.batchtools}~\citep{CRAN:future.batchtools}, which
implements several types of backends on top of the
\CRANpkg{batchtools}~\citep{Lang_etal_2017} package.  Most notably, it
provides backends that resolve futures distributed on high-performance
compute (HPC) environments by submitting the futures as jobs to a job
scheduler, e.g., Slurm, SGE, and Torque/PBS:
\begin{example}
plan(future.batchtools::batchtools_slurm)
plan(future.batchtools::batchtools_sge)
plan(future.batchtools::batchtools_torque)
\end{example}
Yet another example is the \CRANpkg{googleComputeEngineR}
package~\citep{CRAN:googleComputeEngineR}, which provides a ``snow''
cluster type that supports\footnote{It also supports
using \code{parLapply()} functions.}  resolving futures in the cloud
on the Google~Compute~Engine platform.

\section{Implementation}

The future framework is platform-independent and works on all
platforms, including Linux, Solaris, macOS, and MS Windows.  It is
backward compatible with older versions of R back to R 3.1.2 (October
2014).  The core packages \pkg{future},
\CRANpkg{parallelly}~\citep{CRAN:parallelly}, \pkg{globals}, and
\pkg{listenv} are implemented in plain R (without native code) to
maximize cross-platform operability and to keep installation simple.
They are available on CRAN (since 2015).  The \pkg{parallelly} package
implements enhancements to the \pkg{parallel} package originally part
of the \pkg{future} package.  The \CRANpkg{digest}
package~\citep{CRAN:digest} is used to produce universally unique
identifiers (UUIDs).  Development is done toward a public Git
repository hosted at \url{https://github.com/HenrikBengtsson/future}.

\subsection{Validation}
\label{validation}

Since correctness and reproducibility is essential to all data
processing, validation is a top priority and part of the design and
implementation throughout the future ecosystem.  Several types of
testing are performed.

First, all the essential core packages part of the future framework,
\pkg{future}, \pkg{parallelly}, \pkg{globals}, and \pkg{listenv},
implement a rich set of package tests.  These are validated regularly
across the wide range of operating systems (Linux, Solaris, macOS, and
MS Windows) and R versions available on CRAN, on continuous
integration (CI) services (GitHub Actions, Travis CI, and AppVeyor
CI), and on R-hub.

Second, for each new release, these packages undergo full
reverse-package dependency checks
using \pkg{revdepcheck}~\citep{GitHub:revdepcheck}.  As of November 2021,
the \pkg{future} package is tested against 210 direct reverse-package
dependencies available on CRAN and Bioconductor.  These checks are
performed on Linux with both the default settings and when forcing
tests to use multisession workers (SOCK clusters), which further
validates that globals and packages are identified correctly.

Third, a suite of \dfn{Future API conformance tests} available in the
\CRANpkg{future.tests} package~\citep{CRAN:future.tests} validates the
correctness of all future backends.  Any new future backend developed
must pass these tests on complying with the \dfn{Future API}.  By
conforming to this API, the end-user can trust that the backend will
produce the same correct and reproducible results as any other
backend, including the ones that the developer has tested on.  Also,
by making it the responsibility of the backend developer to assert that their
new future backend conforms to the \dfn{Future API}, we relieve other
developers from having to test that their future-based software works
on all backends.  It would be a daunting task for a developer to
validate the correctness of their software with all existing
backends. Even if they would achieve that, there may be additional
third-party future backends that they are not aware of, that they do
not have the possibility to test with, or that yet have not been
developed.

Fourth, since \pkg{foreach} is used by a large number of essential
CRAN packages, it provides an excellent opportunity for supplementary
validation. Specifically, we dynamically tweak the examples of
\pkg{foreach} and popular CRAN packages \CRANpkg{caret},
\CRANpkg{glmnet}, \CRANpkg{NMF}, \CRANpkg{plyr}, and \CRANpkg{TSP} to use
the \CRANpkg{doFuture} adaptor~\citep{CRAN:doFuture}.  This allows us
to run these examples with a variety of future backends to validate
that the examples produce no run-time errors, which indirectly
validates the backends as well as the \dfn{Future API}.  In the past,
these types of tests helped to identify and resolve corner cases where
automatic identification of global variables would fail.  As a side
note, several of these foreach-based examples fail when using a
parallel foreach adaptor because they do not properly export globals
or declare package dependencies.  The exception is when using the
sequential
\dfn{doSEQ} adaptor (default), fork-based ones such as \pkg{doMC}, or
the generic \pkg{doFuture}, which supports any future backend and
relies on the future framework for handling globals and
packages\footnote{There is a plan to update \pkg{foreach} to use the
exact same static-code-analysis method as the \pkg{future} package use
for identifying globals.  As the maintainer of the future framework, I
collaborate with the maintainer of the \pkg{foreach} package to
implement this.}.

Lastly, analogously to the above reverse-dependency checks of each new
release, CRAN and Bioconductor continuously run checks on all these
direct, but also indirect, reverse dependencies, which further
increases the validation of the \dfn{Future API} and the future
ecosystem at large.

\subsection{Known limitations}
\label{known-limitations}

When saving an R object to file or sending it to a parallel worker, R
uses a built-in technique called \dfn{serialization}, which allows a
complex object structure to be sent as a stream of bytes to its
destination, so it later can be reconstructed
via \dfn{unserialization}.  The ability to serialize objects is
fundamental to all parallel processing, the exception being
shared-memory strategies such as forked parallel processing.  For
example, this is how future expressions and variables are sent to
parallel workers and how results are returned.

However, some types of objects are by design bound to the R session
where they are created and cannot be used as-is in other R processes.
One example is R \dfn{connection}s, e.g.,
\begin{example}
con <- file("/path/to/file", open = "wb")
str(con)
#  'file' int 3
#  - attr(*, "conn_id")=<externalptr> 
\end{example}
Any attempt to use a connection in another R process, for instance, by
saving it to file, restarting R, and loading it back in, or by sending
it to a parallel worker, will at best produce a run-time error, and in
the worst case, produce invalid results or, for instance, write to the
wrong file.  These constraints apply to all types of parallelization
frameworks in R, including the future framework.

There are other types of objects that cannot be transferred as-is to
external processes, many from popular third-party packages, e.g.,
database connections of the \CRANpkg{DBI} package, XML documents of
the \CRANpkg{xml2} package, STAN models of the \CRANpkg{stan} package,
and many more\footnote{See \pkg{future} package vignette
\samp{Non-exportable object} for more examples.}.
An indicator of this is when an R object has an \dfn{external
pointer}, which is used for referencing an internal low-level
object. This suggests that the object is bound to the current process
and its lifespan.  Unfortunately, it is not a sufficient indicator
because some objects with external pointers can be exported,
e.g., \CRANpkg{data.table} objects.  This makes it complicated to
automate the detection of non-exportable objects and protect against
using them in parallel processing.  The current best practice is to be
aware of these types of objects and to document new ones when
discovered, which often happens when there is an unexpected run-time
error.  To help troubleshooting, it is possible to configure
the \pkg{future} package to scan for and warn about globals with
external pointers whenever used in a future.

Finally, it is theoretically possible to restructure some of the
``non-exportable'' object types such that they can be used in parallel
processing.  This is discussed further in the `Future work' section.

\subsection{Overhead}
\label{overhead}

With parallel processing comes overhead.  Typically, sources of added
processing time are from spawning new parallel processes,
sending instructions and globals to the workers, querying workers for
results, and receiving results (Figure~\ref{figure:future-4workers}).
Because of this, there is always a trade-off between sequential and
parallel processing, and on how many parallel workers can be used
before the total overhead dominates the benefits.  Whether or not
parallelization is beneficial, and for which parallel backends,
depends on what is being parallelized.

As with other parallel solutions, in the future framework, overhead
differs between parallel backends.  Certain parallel backends, such as
forked processing (``multicore''), are better suited for low-latency
requirements, whereas others, such as distributed processing
(``cluster'' and ``batchtools''), are better suited for
large-throughput requirements.  For example, many fast operations
applied to a single large data frame should probably be parallelized
on the local computer with forked processing, if supported, rather
than being distributed on a compute cluster running in the cloud.  In
contrast, processing hundreds of data files may be completed sooner if
distributed out to multiple computers (with access to the same file
system), for instance, via a job scheduler, rather than being processed
in parallel on the local machine.

Besides the overhead added by the parallel backend, each future,
regardless of backend, has a baseline overhead.  Specifically, there is
a small overhead from the static-code inspection used to identify
global variables, from exception handling needed to capture and relay
errors, and from capturing and relaying standard output and
conditions.  Except for the error-handling overhead, these can all be
avoided via certain \code{future()} arguments, e.g., by manually
specifying globals needed and by disabling the relaying of output and
conditions.

R has several profiling tools that can help identify bottlenecks and
overhead in computational expensive tasks, e.g., \code{system.time()}
of the \pkg{base} package,
\CRANpkg{microbenchmark}~\citep{CRAN:microbenchmark},
\CRANpkg{bench}~\citep{CRAN:bench},
\code{Rprof()} of the \pkg{utils} package,
\CRANpkg{proffer}~\citep{CRAN:proffer}, and
\CRANpkg{profvis}~\citep{CRAN:profvis}.  These tools can also identify
the different sources of overhead in the parallelization framework
itself, including the ones in the future ecosystem.  It is on the
roadmap to make futures collect and report on some of these benchmarks
automatically in order to help developers optimize their code and for
end-users to choose a proper backend.

\section{Results}
\label{results}

The \dfn{Future API} is designed to unify parallel processing in R at
the lowest possible level.  It provides a standard for building
richer, higher-level parallel frontends without having to worry about
and reimplement common, critical tasks such as identifying global
variables and packages, parallel RNG, and relaying of output and
conditions - cumbersome tasks that are often essential to parallel
processing.

Another advantage of the future framework is that new future backends
do not have to implement their versions of these tasks, which not only
lowers the threshold for implementing new backends, but also results
in a consistent behavior throughout the future ecosystem, something
none of the other parallel solutions provide.  This benefits the
developer because they can focus on what to parallelize rather than
how and where. It also benefits the end-user, who will have more
alternatives to how and where parallelization will take place.  For
instance, the developer might have local parallelization in mind during
the development phase due to their work-environment constraints,
whereas the end-user might be interested in parallelizing out to a
cloud computing service.  One may say that code using futures scales far
without the developer's attention.  Moreover, code using futures for
parallelization will be able to take advantage of new backends that
may be developed several years from now.

Directly related to the separation of code and backends, 
end-users and developers no longer need to rely on other package
maintainers to update their code to take advantage of any new types of
computational resources; updates that otherwise require adding another
argument and conditional statement.  One example of this
was \pkg{future.batchtools}' predecessor, \CRANpkg{future.BatchJobs}
(legacy, CRAN, archived), which was straightforward to implement on
top of \CRANpkg{BatchJobs}~\citep{Bischl_etal_2015} as soon as
the \dfn{Future API} was available.  With zero modifications, code
that previously only parallelized on the local computer could
suddenly parallelize across thousands of cores on high-performance
compute (HPC) clusters via the job scheduler.  All it took was to
change the \code{plan()}.

Because the future ecosystem is at its core designed to give
consistent results across all sequential and parallel backends, it is
straightforward to update, or port, an existing, sequential,
map-reduce framework such that it can run in parallel. Not having to
worry about low-level parallelization code, which otherwise risks
blurring the objectives, lowers the threshold for designing and
implementing new parallel map-reduce APIs.  There are several examples
of how fairly straightforward it is to implement higher-level parallel
APIs on top of the \dfn{Future API}.  The \pkg{future.apply}
package~\citep{CRAN:future.apply}, implements futurized variants of
R's apply functions found in the \pkg{base} package,
e.g., \code{future\_apply()} and \code{future\_lapply()} are plug-in
replacements for \code{apply()} and \code{lapply()}.  The \pkg{furrr}
package~\citep{CRAN:furrr} implements futurized variants of the
different map-reduce functions found in the \CRANpkg{purrr}
package~\citep{CRAN:purrr}, e.g., \code{future\_map()} is as plug-in
replacement for \code{map()}.  The \pkg{doFuture} package implements a
generic \pkg{foreach} adaptor
for \code{y\,<-\,foreach(...)\,\%dopar\%\,\{\,...\,\}} that we can use
with any future backend.  Because
the \BIOpkg{BiocParallel}~\citep{Bioc:BiocParallel} package, part of
the Bioconductor Project, supports foreach as its backend, its
functions such as \code{bplapply()} and \code{bpvec()} can also
parallelize using \emph{any type of future backend} via
\pkg{doFuture}.

By lowering the barrier for implementing futurized variants of popular
map-reduce APIs, developers and end-users are allowed to stay with
their favorite coding style while still taking full advantage of the
future framework.

The \dfn{Future API} also addresses the lock-in-versus-portability
problem mentioned in the introduction; the risk that package
developers on Unix-like systems would only support multicore
parallelization methods because ``\code{mclapply()} just works'' is
significantly lower using futures.  Similarly, the most common way to
parallelize code is to use multiple cores on the local machine.
Because it is less common to have access to multiple machines, this
often prevents developers from considering any other types of
parallelization, with the risk of locking in end-users with other
types of resources to only use a single machine.  Hence, the chance for a
package to support multi-host parallelization, including in the cloud
and HPC environments, increases when using futures.

The burden on package developers to test and validate their parallel
code is significant when using traditional parallelization frameworks,
especially when attempting to support multiple variants.  In contrast,
when using futures, the cost of developing, testing, and maintaining
parallel code is lower - often not much more than maintaining
sequential code.  This is possible because of the simplicity of the
\dfn{Future API} and the fact that the orchestration of futures is
predominantly done by the \pkg{future} package.  Therefore, by
implementing rigorous tests for the future framework and the different
backend packages, the need for performing complementary tests in
packages that make use of futures is much smaller.  Tests for future
backend packages, as well as the \dfn{Future API}, are provided by
the \pkg{future.tests} package, which lowers the risk for a backend
not being sufficiently tested.

The built-in protection against nested parallelism by mistake, and the
agility of system settings of \code{availableCores()}, makes parallel
code that uses futures to play nicely on multi-tenant systems.  It
respects all known R options and environment variables that specify,
or otherwise limit the number of parallel workers allowed.
See \code{help("availableCores",\,package\,=\,"parallelly")} for
details.  In contrast, it is, unfortunately, very common to find
parallel code that uses \code{parallel::detectCores()} as the default
number of workers in other parallel frameworks.  Defaulting to using
all available cores this way often wreak havoc on multi-tenant compute
systems by overusing already consumed CPU resources, sometimes
bringing the system to a halt due to too much context switching and
memory use.  Unfortunately, this often results in a negative
performance on also other users' processes, and system administrators
have to spend time tracking down the root cause of such poorly
performing compute hosts.

\subsection{Use of the future framework on CRAN and Bioconductor}
\label{use-of-the-future-framework}

The \pkg{future} package was released on CRAN in 2015. The uptake has
grown steadily ever since.  As of November 2021, \pkg{future} is among
the top-1.1\% most downloaded package on CRAN\footnote{The ranks are
robust estimates based on the average median weekly download counts
from the RStudio CRAN mirror during four weeks.}, and there are 210
packages on CRAN and Bioconductor that directly depend on it.  For
map-reduce parallelization packages \pkg{future.apply} (top-1.3\% most
downloaded) and \pkg{furrr} (top-1.8\%), the corresponding number of
packages are 87 and 58, respectively.

Besides supporting these traditional parallelization methods, the
future framework is also used as an infrastructure elsewhere.  For
example, the workflow package \CRANpkg{targets}~\citep{Landau_2021},
and its predecessor \CRANpkg{drake}~\citep{Landau_2018}, implements
``a pipeline toolkit for reproducible computation at scale''.  They
work by defining make-like targets and dependencies that can be
resolved in parallel using any type of future backend.  Another
prominent example is the \CRANpkg{shiny} package~\citep{CRAN:shiny},
which implements support for \dfn{asynchronous processing} in Shiny
applications via futures.  Asynchronous processing helps to avoid
long-running tasks from blocking the user interface.  Similarly,
the \CRANpkg{plumber} package~\citep{CRAN:plumber}, which
automatically generates and serves HTTP API from R functions, uses
futures to serve asynchronous web APIs and process tasks in parallel.

\subsection{Other uses of futures}
\label{other-usages}

In \citet{HewittBaker_1977}, the authors propose the
\code{(EITHER\ ...)} construct that ``evaluates the expressions in
parallel and return the value of 'the first one that finishes'.''  A
corresponding R construct could be \code{future\_either(...)} that
evaluates R expressions concurrently via futures and returns the value
of the first resolved one ignoring the others, e.g.,
\begin{example}
y <- future_either(
  sort.int(x, method = "shell"),
  sort.int(x, method = "quick"),
  sort.int(x, method = "radix")
)
\end{example}

We may also use futures in cases that do not require parallel
processing per se.  Indeed, the \dfn{Future API} strives to make no
assumptions about futures being resolved via parallel or distributed
processing.  One example is where a particular expression can only be
resolved in a legacy version of R, on another operating system than
where the main R session runs, or in an environment that meet specific
requirements, e.g., large amounts of memory, fast local disks, or
access to a certain genomic database.  Another example of a
resource-specific backend is the \CRANpkg{civis}
package~\citep{CRAN:civis}, which uses futures to provide an R client
for the commercial Civis Platform.

We can also use futures to evaluate non-trustworthy R expressions in a
sandboxed R environment that is, for instance, locked down in a
virtual machine, or in a Linux container, such as
Singularity~\citep{Kurtzer2017} or Docker~\citep{Merkel2014}, without
access to the host machine and its file system and network.

\section{Future work}
\label{future-work}

Although they are not part of the core future framework, future-based
map-reduce packages \pkg{future.apply}, \pkg{furrr}, \pkg{doFuture},
and the like, play an essential role in how developers and end-users
interact with futures.  A key feature of these packages is ``load
balancing'', which helps reduce the overall overhead that comes from
setting up futures and spawning them on parallel workers and
collecting their results.  They achieve this by partitioning the
elements to iterated over into equally sized chunks, typically so that
there is one chunk per worker, which in turn results in one future per
chunk and hence one future per worker.  In contrast, without load
balancing, each element is processed by one future resulting in more
overhead, especially when there are many elements to iterate over.
Each of these packages has its own implementation of load balancing,
despite often using exactly the same algorithm.  If there is an
improvement or a bug fix to one, the maintainers of the others need to
update their code too.  The same is true for how they orchestrate
globals and parallel RNG.  To improve on this situation and to further
harmonize the behavior of futures in these packages, a new helper
package \pkg{future.mapreduce} that implements these common tasks will
be introduced, relieving these packages from those tasks.  This will
also have the advantage of making it even easier to implement other
types of map-reduce APIs on top of futures.

Having said this, in a longer perspective, it might be possible to
remove the need for these future-based map-reduce APIs, which
essentially are thin wrappers ported from their counterpart
map-reduce APIs.  This would require internal refactoring of the core
future framework, but it can likely be done while preserving full
backward compatibility with the current \dfn{Future API}.
For clarification, consider the following \code{lapply()} construct
that evaluates \code{slow\_fcn(x)} for ten elements, each resolved via
a unique \dfn{lazy} future:
\begin{example}
xs <- 1:10
fs <- lapply(xs, function(x) future({
  slow_fcn(x)
}, lazy = TRUE))
\end{example}
A lazy future defers the evaluation of its expression until we
use \code{resolved()} to query if it is resolved or until we
use \code{value()} to collect its value\footnote{Although a lazy
future defers the evaluation to a later time, contrary to
R's \dfn{lazy evaluation} and \dfn{promises}, a future records all
dependent variables (``globals'') when it is created, which means it
will resolve to the same value even if those globals change after the
future was created and before it was resolved.  This also means that
lazy and eager futures give the same value.}.  Since neither has been
called above, these futures are still dormant, regardless of future
backend used.  Next, assume that there are two parallel workers and
imagine that we have a function \code{merge()} to merge futures.  This
would allow us to partition ten futures into only two futures, one per
worker, and then collect their values:
\begin{example}
f1 <- merge(fs[1:5])
f2 <- merge(fs[6:10])
vs <- c(value(f1), value(f2))
\end{example}
We can simplify this further by encapsulating the above in the S3
method \code{value()} for \dfn{list}s:
\begin{example}
vs <- value(fs)
\end{example}
We can mitigate the verbosity in the setup of futures with a helper
function or syntax sugar.  More importantly, this would make it
possible to use futures in map-reduce APIs without the need for a
counterpart parallel implementation.  It would also lower the
threshold further for adding a thin layer of support for
futures \emph{within} existing map-reduce APIs, especially since the
design of the future framework keeps the added maintenance burden to a
minimum.

A frequently requested feature is to support \emph{suspending} running
futures, particularly when their runtimes are large.  For example,
above \code{future\_either()} function could benefit from
a \code{suspend()} function to terminate futures no longer needed.
Since not all backends may support
this, extra care needs to be taken when introducing this feature to
the future framework.  A related feature request is the possibility
to \emph{restart} a future that failed due to, for instance, a crashed
worker or a partial power failure on a compute cluster,
e.g., \code{restart(f)}.  Combined with R's condition handling
framework, higher-level APIs can then take on the role of retrying to
resolve failed futures,
e.g., \code{retry(\{\,...\,\},\,times\,=\,3,\,on\,=\,"FutureError")}.

Implementing support for suspending and restarting futures will
indirectly add support for serializing futures themselves, which is
only partially supported in the current implementation.  Being able to
serialize futures opens up further possibilities such as saving
futures to be processed at a later time, in another context, or
transferring them to a job queue that, in turn, distributes them to
appropriate compute resources.

The problem of not being able to export all types of objects as-is in
parallel processing can be a blocker.  It turns out that for a subset
of these, we could use \dfn{marshaling} to encode them before
serializing them such that a working clone can be reconstructed after
unserializing and \dfn{unmarshaling}.  As an example, a read-only file
connection can be marshaled by recording its filename and file position
so that the parallel worker could open its own read-only connection
for the same file at the same position.  Marshaling is a rarely used
concept in R, possibly because there is no standard convention for
package developers to rely on. Ideally, such a mechanism would allow
package developers to register custom \code{marshal()}
and \code{unmarshal()} methods for their data types, making them
automatically applicable in parallelization without prior knowledge of
what objects being transferred.

Other than setting the backend via \code{plan()}, it is not possible
to direct a particular future to a specific backend type based on the
needs of the future.  To support this, we have to add options to
declare what \dfn{resources} are needed to resolve particular future.
For instance,
\begin{example}
f <- future({ ... }, resources = c("r:3.2.*", "mount:/data", "!fork"))
\end{example}
could be one way to specify that this future has to be resolved with R
3.2 on a machine with a \code{/data} mount point and that forked
parallel processing must not be used.  Some resources may be implicit
based on exported globals, e.g., a specific file required when
exporting a file connection via marshaling.

All the above is on the roadmap for the future framework.

\section{Summary}

The \pkg{future} package is a lightweight R package that provides an
alternative approach for parallel processing in R.  It implements
the \dfn{Future API}, which comprises three basic functions, from which
richer, higher-level APIs for parallel processing can be constructed.
Several of these higher-level APIs mimic counterpart map-reduce APIs
closely, allowing developers to stay with their favorite coding
style for their parallel needs.  The future framework is designed
so that the developer does not have to worry about common, critical
tasks such as exporting globals to workers, using proper parallel RNG,
and taking care of output, messages, warnings, and errors.  This design
lowers the barriers to reimplement existing algorithms and methods in
parallel while avoiding increasing the maintenance burden.  When
using futures, the end-user controls which parallel backend is used,
while the developer controls what to parallelize.  This is possible
because all future backends have been validated to conform to
the \dfn{Future API} specifications, ensuring that futures produce the
same results regardless of how and where they are processed.

\section{Acknowledgments}
\label{acknowledgements}
 
I am grateful to all users and developers who have contributed to the
future framework with questions and answers, feature requests, bug
reports, and interesting and fruitful discussions.  I am thankful to
the reviewers, the editor, and everyone else who gave comments and
suggestions helping to improve this article.  The development of
the \dfn{Future API Specifications and Conformance} test suite in
the \pkg{future.tests} package was supported by the R~Consortium
through its Infrastructure Steering Committee (ISC) grant program.

\bibliography{bengtsson-future}

\address{%
Henrik Bengtsson\\
Department of Epidemiology and Biostatistics,\\
University of California, San Francisco\\
San Francisco, CA\\
United States\\
\email{henrik.bengtsson@ucsf.edu}
}
\end{article}

\pagestyle{plain}
\end{document}